\documentclass[prb,twocolumn,english]{revtex4}
\usepackage{amsmath}
\usepackage{graphicx}
\usepackage{epstopdf}
\begin{document}

\title{Analysis of the quantum-classical Liouville equation in the mapping basis}

\author{Ali Nassimi$^1$, Sara Bonella$^2$ and Raymond Kapral$^1$}

\affiliation{$^1$Chemical Physics Theory Group, Department of Chemistry, University
of Toronto, Toronto, Ontario M5S 3H6, Canada\\
$^2$Dipartimento di Fisica and CNISM Unit 1, Universit\`a La Sapienza, P.le A. Moro 2, 00185 Rome}

\email{anassimi@chem.utoronto.ca}

\email{rkapral@chem.utoronto.ca}

\email{sara.bonella@roma1.infn.it}

\begin{abstract}
The quantum-classical Liouville equation provides a description of the dynamics of a quantum subsystem coupled to a classical environment. 
Representing this equation in the mapping basis leads to a continuous description of discrete quantum states of the subsystem and may provide an alternate route to the construction of simulation schemes. 
In the mapping basis the quantum-classical Liouville equation consists of a Poisson bracket contribution and a more complex term. 
By transforming the evolution equation, term-by-term, back to the subsystem basis, the complex term (excess coupling term) is identified as being due to a fraction of the back reaction of the quantum subsystem on its environment. 
A simple approximation to quantum-classical Liouville dynamics in the mapping basis is obtained by retaining only the Poisson bracket contribution. 
This approximate mapping form of the quantum-classical Liouville equation can be simulated easily by Newtonian trajectories. 
We provide an analysis of the effects of neglecting the presence of the excess coupling term on the expectation values of various types of observables. 
Calculations are carried out on nonadiabatic population and quantum coherence dynamics for curve crossing models. 
For these observables, the effects of the excess coupling term enter indirectly in the computation and good estimates are obtained with the simplified propagation.
\end{abstract}

\maketitle

\section{introduction}

The construction of algorithms for simulating the quantum dynamics of an arbitrary
many-body system is a long standing problem. Since the computational cost of a quantum simulation scales exponentially with the system size, a full quantum calculation is impractical except for small systems. This fact has
prompted the construction of a variety of approximate methods for the simulation of quantum dynamics. One class of approximate schemes singles out a portion of the system for a full quantum treatment while
its environment---the rest of the system---is treated classically. A number of
such mixed quantum-classical schemes have been proposed and references to this literature can be found in reviews on this topic~\cite{gdb,tullyhop,herman,truhlarrev,Rayrev06}.

We focus on one scheme of this type based on the quantum-classical Liouville equation
(QCLE),
\begin{eqnarray} \label{quantum-classical}
\frac{\partial}{\partial t}\hat{\rho}_W(X,t)&=& -\frac{i}{\hbar}[\hat{H}_W,\hat{\rho}_W(t)] \\
&& +\frac{1}{2} \big(\{\hat{H}_W,\hat{\rho}_W(t)\}-\{\hat{\rho}_W(t),\hat{H}_W\}\big),\nonumber
\end{eqnarray}
where $[\cdot,\cdot]$ is the commutator and $\{\cdot,\cdot\}$ is the Poisson bracket. Here $X=(X_1,X_2, \dots, X_{N_e})=(R,P)=(R_1,R_2, \dots, R_{N_e}, P_1,P_2, \dots, P_{N_e})$ are the positions and momenta of the $2{N_e}$ environmental degrees of freedom and the index W stands for a partial Wigner transform~\cite{imre,hillery} over the environmental degrees of freedom, so that the density matrix, $\hat{\rho}_W(X)$, and Hamiltonian, $\hat{H}_W(X)$, are operators in the Hilbert space of the subsystem while functions of the phase space variables for the bath degrees of freedom. (The dependence on the phase space coordinates will be omitted when confusion is unlikely to arise.) The nature of QCL dynamics and the statistical mechanics of systems following this dynamics have been described~\cite{steve}. For a review with references to the literature on this topic, see Ref.~[\onlinecite{Rayrev06}].

While the QCLE has a number of attractive features and has been shown to provide an
accurate description of the dynamics in many instances~\cite{donmar,donzhenmar,marfan,riga06,santer01,horenko02,mukamel94,toutounji,gabray,Hanna:Kapral,Hankap,Kim:Hanna,hanna08}, it is difficult to solve.
A direct numerical integration of Eq.~(\ref{quantum-classical}) requires very fine
spatial grids and can be used for the study of small systems~\cite{marfan}. The QCLE can be cast in any basis which spans the Hilbert space of the quantum subsystem~\cite{RAR}. When written in the quantum subsystem
basis, the QCLE can be solved using a trajectory-based algorithm where trajectories
are not independent of one another~\cite{donmar,donzhenmar}.
Representation of QCLE in the adiabatic basis gives a more intuitive picture in terms of
classical trajectories moving on single or mean Born-Oppenheimer surfaces~\cite{ray99,Sergi:MacKernan,mck}.
Also, diagonalizing the Hellman-Feynman force derived from the Born-Oppenheimer potential
leads to the force basis which yields yet a different route to simulation of the QCLE~\cite{ws00,ws02}.

Another representation of QCL dynamics is in terms of the mapping basis~\cite{hyo:ray:ali,Ali}. This basis, which provides a description of a
discrete quantum system in continuous variables, has been used in a number of different
applications to quantum dynamical problems~\cite{sch,holpri40,Mill:Mcur,sun98,Miller01,stotho,ThoSto99,MulSto98,StoTho97,bonella03,bonella05,Dunkel08}.
When the QCLE is expressed in this basis one obtains an evolution equation where the evolution operator consists of a Poisson bracket contribution which can be solved by characteristics, and a more complex term that involves specific correlations of the quantum subsystem and classical environment. Thus far, simulations of the QCLE in the mapping basis have neglected this more complex contribution. In this paper we provide an analysis of the mapping form of the QCLE that elucidates the nature of this neglected term. By transforming the QCLE in the mapping basis back to the subsystem basis, where an interpretation of the physical meaning of the different terms is easier, we show that the neglected term (called the excess coupling term) accounts for a fraction of back reaction of the quantum subsystem on its environment. We also describe how the contribution of the excess coupling term to the average values of certain observables can be estimated numerically.

The outline of the paper is as follows: In Sec.~\ref{Mapping} we summarize the equations to be analyzed and introduce some key definitions, while in Sec.~\ref{transformation} we perform the term-by-term analysis of the mapping equation based on the back transformation to the subsystem basis. Section~\ref{sec:approx} presents an analysis of the approximate form of the mapping QCLE that includes only the Poisson bracket term, called the Poisson bracket mapping equation (PBME). This section also describes how neglecting the excess coupling term affects average values of different types of observables. In Sec.~\ref{Numerical investigations}, the performance of the PBME is tested on two often-studied curve crossing models, and the effects of the excess coupling term in the evolution operator are discussed. The last section contains the conclusions of our investigation.

\section{Subsystem and Mapping Basis Representations} \label{Mapping}

For a system with partially Wigner transformed Hamiltonian $\hat H_W = \frac{P^2}{2M}+ \frac{\hat p^2}{2m}+\hat V_s(\hat r)+V_e(R)+\hat V_c(R,\hat r)$, where $V_e$, $\hat V_s$ and $\hat V_c$ are, respectively, the environment,
subsystem and coupling potentials, $R$ and $P$ are again the $N_e$-dimensional coordinates and momenta of particles of the environment with mass $M$, and ${\hat r}=({\hat r}_1,{\hat r}_2,\dots,{\hat r}_{N_s})$ and  $\hat{p}=(\hat{p}_1,\hat{p}_2,\dots,\hat{p}_{N_s})$ are the $N_s$ coordinate and momentum operators of the particles of the quantum subsystem with mass $m$, the QCLE in the subsystem basis takes the form~\cite{ray99},
\begin{eqnarray}
\label{QCLE in SSB}
&&\frac{\partial }{\partial t}\rho_W^{\alpha\alpha'}(X,t) = -i\tilde \omega_{\alpha\alpha'}\rho_W^{\alpha\alpha'}(t) \nonumber \\
&& \qquad + \frac{i}{\hbar} \big(\rho_W^{\alpha\alpha''}(t)V_c^{\alpha''\alpha'} - V_c^{\alpha\alpha''}\rho_W^{\alpha''\alpha'}(t)\big) \nonumber \\
&& \qquad + \Big( {\partial V_e\over \partial R}\cdot \frac{\partial}{\partial P}-\frac{P}{M} \cdot \frac{\partial}{\partial R}\Big)\rho_W^{\alpha\alpha'}(t)
\nonumber \\
&& \qquad + \frac{1}{2} \Big(\frac{\partial V_c^{\alpha\alpha''}}{\partial R}\cdot \frac{\partial \rho_W^{\alpha''\alpha'}(t)}{\partial P} + \frac{\partial \rho_W^{\alpha\alpha''}(t)}{\partial P} \cdot \frac{\partial V_c^{\alpha''\alpha'}}{\partial R}\Big)\nonumber \\
&&\quad = -i\tilde{\mathcal L}_{\alpha \alpha', \mu \mu'} \rho^{\mu \mu'}_W(X,t).
\end{eqnarray}
Here and in the following we use the Einstein summation convention where repeated indices are summed.
The subsystem basis is defined by the eigenvalue problem, $\hat{h}_s|\alpha\rangle = \epsilon_{\alpha}|\alpha \rangle$, where $\hat{h}_s=\frac{\hat p^2}{2m}+\hat V_s$. In Eq.~(\ref{QCLE in SSB}),
$\tilde \omega_{\alpha\alpha'} = (\epsilon_\alpha - \epsilon_{\alpha'})/\hbar$ and $V_c^{\alpha\alpha'} = \langle \alpha|\hat V_c|\alpha'\rangle $. The last line in Eq.~(\ref{QCLE in SSB}) defines $\tilde{\mathcal L}_{\alpha \alpha', \mu \mu'}$, the QCL operator in the subsystem basis.

Starting from this subsystem representation of the QCLE, we may transform it to the mapping basis in the following way. Suppose that there are $N$ subsystem quantum states $|\lambda \rangle$, $\lambda =1, \dots, N$. These subsystem states may be mapped onto harmonic oscillator states via the transformation, $|\lambda \rangle \rightarrow |m_\lambda \rangle = |0_1,\cdots,1_\lambda,\cdots,0_N \rangle$, where
\begin{eqnarray} \label{map}
\langle q|m_{\lambda}\rangle &=& \langle q_{1},q_{2},\cdots,q_{N}|0_{1},\cdots,1_{\lambda},\cdots,0_{N}\rangle \nonumber \\
&=&\phi_{0}(q_{1})\cdots\phi_{0}(q_{\lambda-1})\phi_{1}(q_{\lambda})\cdots\phi_{0}(q_{N}),
\end{eqnarray}
with $\phi_{0}$ and $\phi_{1}$, respectively, being the ground and the first excited state wave functions of a harmonic
oscillator. The creation and annihilation operators on the mapping states, $\hat{a}_{\lambda} = (\hat{q}_{\lambda}+i\hat{p}_{\lambda})/\sqrt{2\hbar}$ and $\hat{a}_{\lambda}^{\dag}=(\hat{q}_{\lambda}-i\hat{p}_{\lambda})/\sqrt{2\hbar}$,
are used to define a mapping of operators as~\cite{footnote-units}
\begin{equation}\label{eq:map-operator}
\hat A = A^{\lambda \lambda'} |\lambda\rangle \langle\lambda'| \rightarrow \hat A_m = A^{\lambda \lambda'} \hat{a}_{\lambda}^\dag \hat{a}_{\lambda'}.
\end{equation}
For example, the partially Wigner transformed mapping form of the system Hamiltonian is
\begin{eqnarray} \label{hamiltonian2}
\hat{H}_m &=& \frac{P^2}{2M} + V_e(R) + \frac{1}{2\hbar}h^{\lambda\lambda'}(R) \big(\hat q_{\lambda} \hat q_{\lambda'} + \hat p_{\lambda} \hat p_{\lambda'} - \hbar\delta_{\lambda\lambda'}\big)\nonumber \\
&\equiv&\frac{P^2}{2M} + V_e(R) + \hat{h}_m(R).
\end{eqnarray}
Here we used the expression for the creation and annihilation operators in terms of the coordinates and momenta of the harmonic oscillators, together with the appropriate commutation relationship. Also, $h^{\lambda \lambda'}(R) = \langle \lambda |\hat{h}| \lambda' \rangle$  with $\hat{h}=\hat{h}_s +\hat{V}_c(R,\hat{r})$ so that, more explicitly,
\begin{equation} \label{hamiltonian-matrix}
h^{\lambda \lambda'}(R)=\epsilon_\lambda \delta_{\lambda \lambda'} +V_c^{\lambda \lambda'}(R).
\end{equation}
In writing Eq.~(\ref{hamiltonian2}), we assumed that $h^{\lambda \lambda'} = h^{\lambda' \lambda}$. Following this prescription, after a further Wigner transformation~\cite{imre} over the mapping variables,
\begin{equation}\label{eq:map-wigner-rho}
{\rho}_m(X,x)=\frac{1}{(2\pi \hbar)^{N}}\int dz\; e^{ip\cdot z/\hbar} \langle r- \frac{z}{2}|\hat{\rho}_m(X)|r+ \frac{z}{2}\rangle,
\end{equation}
where $x=(x_1,x_2, \dots, x_N)=(r,p)=(r_1,r_2, \dots, r_N,p_1,p_2, \dots, p_N)$, the QCLE takes the form~\cite{hyo:ray:ali,Ali}
\begin{eqnarray} \label{ultimate_expression_for_QCL1}
&&\frac{\partial}{\partial t} \rho_m(X,x,t) = - \frac{h^{\lambda\lambda'}}{\hbar}\left(p_{\lambda'} \frac{\partial}{\partial r_{\lambda}}-r_{\lambda'} \frac{\partial}{\partial p_{\lambda}} \right) \rho_m(t) \\
&& \qquad \qquad + \Big(\frac{\partial H_m}{\partial R}\cdot  \frac{\partial }{\partial P}
-\frac{P}{M}\cdot \frac{\partial }{\partial R} \Big) \rho_m(t) \nonumber \\
&& \qquad \qquad - \frac{\hbar}{8} \left[\frac{\partial h^{\lambda\lambda'}}{\partial R}\Big(\frac{\partial^2}{\partial r_{\lambda} \partial r_{\lambda'}}+ \frac{\partial^2}{\partial p_{\lambda} \partial p_{\lambda'}}\Big)\cdot \frac{\partial \rho_m(t)}{\partial P}\right].\nonumber
\end{eqnarray}
Note that the Wigner transform variables $r$, $p$ and $z$ have the same dimension as the number of quantum states, $N$.
The Wigner transform of $\hat{H}_m$ over mapping variables is
\begin{eqnarray} \label{h_m-def}
&&{H}_m(X,r,p)=\int dz\; e^{ip\cdot z/\hbar} \langle r- \frac{z}{2}|\hat{H}_m|r+ \frac{z}{2}\rangle \nonumber\\
&&= \frac{P^2}{2M} + V_e(R) + \frac{1}{2\hbar}h^{\lambda\lambda'}(R) \big( r_{\lambda}  r_{\lambda'} +  p_{\lambda} p_{\lambda'} - \hbar\delta_{\lambda\lambda'}\big)\nonumber \\
&&\equiv \frac{P^2}{2M} + V_e(R) + {h}_m(R).
\end{eqnarray}

We can also write the mapping form of the QCLE as
\begin{eqnarray} \label{map-QCL1}
&&\frac{\partial}{\partial t} \rho_m(X,x,t) = \{H_m,\rho_m(t)\}_{X,x} \\
&& \qquad \qquad - \frac{\hbar}{8} \left[\frac{\partial h^{\lambda\lambda'}}{\partial R}\Big(\frac{\partial^2}{\partial r_{\lambda} \partial r_{\lambda'}}+ \frac{\partial^2}{\partial p_{\lambda} \partial p_{\lambda'}}\Big)\cdot \frac{\partial \rho_m(t)}{\partial P}\right].\nonumber
\end{eqnarray}
In Eq.~(\ref{map-QCL1}), $\{H_m,\rho_m\}_{X,x}$ denotes a Poisson bracket with respect to both the bath $X$ and mapping $x$ variables. Neglecting the last term of the dynamics in Eq.~(\ref{map-QCL1}) yields a Hamiltonian system of equations which can be easily solved using Newtonian trajectories.
The last term has a complex form involving both bath and mapping differential operators which make its interpretation difficult and precludes the implementation of simple
algorithms for the simulation of the full mapping form of the QCLE.

\section{Transforming mapping dynamics back to the subsystem basis} \label{transformation}

In order to understand the nature of the last term in the mapping QCLE, we shall transform each term in this equation back to the quantum subsystem basis. Naturally, combining all the back-transformed terms, the original subsystem basis equation (Eq.~(\ref{QCLE in SSB})) is recovered; however, the term-by-term transformation highlights how each contribution in the mapping representation of the QCL operator is associated with a specific contribution in the subsystem basis form of the operator. This procedure leads to a simple physical interpretation of the last term in Eq.~(\ref{map-QCL1}).

We begin by recalling the expression for the Wigner transformed density matrix in the mapping basis.
Using the definition of a mapping operator in Eq.~(\ref{eq:map-operator}) and the expression for the Wigner transform of the density operator in Eq.~(\ref{eq:map-wigner-rho}), one has
\begin{eqnarray}
\label{SMtransform}
&&\rho_{m}(X,x) = \frac{1}{(2\pi\hbar)^N}\int dz e^{ip\cdot z/\hbar} \langle r-\frac{z}{2}|\hat \rho_m(X)|r+\frac{z}{2}\rangle \nonumber \\
&&\quad = \frac{1}{(2\pi\hbar)^N} \int dz e^{ip\cdot z/\hbar} \rho^{\lambda\lambda'}_W(X) \langle r-\frac{z}{2}|\hat a_{\lambda}^\dag \hat a_{\lambda'}|r+\frac{z}{2}\rangle \nonumber \\
&&\quad =\rho^{\lambda\lambda'}_W(X) c_{\lambda\lambda'}(x),
\end{eqnarray}
where
\begin{eqnarray}
\label{FD}
&&c_{\lambda\lambda'}(x)
= \frac{1}{2\hbar(2\pi\hbar)^N} \\
&&\quad \times \Big[r_\lambda r_{\lambda'} + i(r_\lambda p_{\lambda'} - r_{\lambda'} p_{\lambda}) + p_{\lambda}p_{\lambda'} - \hbar\delta_{\lambda\lambda'} \Big]. \nonumber
\end{eqnarray}
Details of the calculation of $c_{\lambda\lambda'}(x)$ are given in the Appendix.

Next, given $\rho_{m}(X,x)$ we consider how to recover the expression for $\rho^{\lambda \lambda'}_W(X)$. The mapping relationship in Eq.~(\ref{eq:map-operator}) ensures that the matrix elements of an operator are the same in the subsystem and mapping bases, thus,
\begin{equation}\label{eq:matrix-id}
\rho^{\lambda \lambda'}_W(X)\equiv \langle \lambda|\hat{\rho}_W|\lambda' \rangle  = \langle m_{\lambda}|\hat \rho_m|m_{\lambda'}\rangle.
\end{equation}
Inserting resolutions of the identity in the coordinate representation, the last term in the equality above can be written as
\begin{eqnarray}
&&
\langle m_{\lambda}|\hat \rho_m|m_{\lambda'}\rangle = \int dydy'\; \langle m_{\lambda}|y\rangle\langle y|\hat \rho_m|y'\rangle\langle y'|m_{\lambda'}\rangle  \\
&& \quad = \int drdz\; \langle m_{\lambda}|r-\frac{z}{2}\rangle\langle r-\frac{z}{2}|\hat \rho_m|r+\frac{z}{2}\rangle\langle r+\frac{z}{2}|m_{\lambda'}\rangle \nonumber
\end{eqnarray}
Mean $r=(y+y')/2$ and difference $z=y'-y$ coordinates were introduced in the last line to pave the way for the introduction of the Wigner transform of the density operator in the mapping coordinates. The off-diagonal element of the operator in the equation above can in fact be expressed as  $\langle r-\frac{z}{2}|\hat \rho_m|r+\frac{z}{2}\rangle=\int dp e^{-ip\cdot z/\hbar}\rho_{m}(X,x)$ (the inverse transform of the expression in Eq.~(\ref{eq:map-wigner-rho})). Combining this expression with the identity in Eq.~(\ref{eq:matrix-id}), we get
\begin{eqnarray}
&& \rho^{\lambda \lambda'}_W(X) = \label{MStransform} \\
&&\quad \int drdzdp\; e^{-ip\cdot z/\hbar} \rho_{m}(X,x) \langle m_{\lambda}|r-\frac{z}{2}\rangle \langle r+\frac{z}{2}|m_{\lambda'}\rangle \nonumber
\end{eqnarray}
In the Appendix we show that the integral over $z$ in the equation above can be performed analytically to obtain
\begin{eqnarray}\label{MStransform}
\rho^{\lambda \lambda'}_W(X) = \int dx\; \rho_{m}(X,x) g_{\lambda \lambda'}(x),
\end{eqnarray}
where
\begin{eqnarray} \label{GD}
&&g_{\lambda \lambda'}(x) = \frac{2^{N+1}}{\hbar} e^{-x^2/\hbar} \\
&& \qquad \times \Big[r_\lambda r_{\lambda'} - i(r_\lambda p_{\lambda'} - r_{\lambda'} p_{\lambda}) + p_{\lambda}p_{\lambda'} - \frac{\hbar}{2}\delta_{\lambda\lambda'} \Big]. \nonumber
\end{eqnarray}
(In the notation of this paper $x^2=x \cdot x= r^2 +p^2= \sum_\lambda(r_\lambda^2+p_\lambda^2)$. For clarity, in some places we shall use the more expanded forms of this condensed notation.) Relationships analogous to those in Eqs.~(\ref{SMtransform}) and (\ref{MStransform}) hold for the representation of any operator in the two bases, except for the fact that the multiplicative prefactors in Eqs.~(\ref{FD}) and (\ref{GD}) are interchanged.

These results can be used to establish the relationship between the mapping and the subsystem bases. To set the stage for our analysis, we multiply each term of Eq.~(\ref{map-QCL1}) by $g_{\alpha\alpha'}(x)$ and integrate over the mapping coordinates to obtain
\begin{eqnarray} \label{map-QCL1-back-0}
&&\frac{\partial}{\partial t} \rho_W^{\alpha \alpha'}(X,t) = \int dx\; g_{\alpha \alpha'}(x)\{H_m,\rho_m(t)\}_{X,x} \nonumber \\
&&\quad   - \frac{\hbar}{8} \int dx\; g_{\alpha \alpha'}(x)\left[\frac{\partial h^{\lambda\lambda'}}{\partial R}\Big(\frac{\partial^2}{\partial r_{\lambda} \partial r_{\lambda'}}+ \frac{\partial^2}{\partial p_{\lambda} \partial p_{\lambda'}}\Big)\cdot \frac{\partial }{\partial P}\right]\nonumber \\
&& \quad \times \rho_m(t).
\end{eqnarray}
In the expression above, the integrand contains $\rho_m(X,x,t)$. We can use Eq.~(\ref{SMtransform}) to express this quantity in terms of the density matrix in the subsystem basis, $\rho^{\mu \mu'}_W(t)$, to get
\begin{eqnarray} \label{map-QCL1-back}
&&\frac{\partial}{\partial t} \rho_W^{\alpha \alpha'}(X,t) = \int dx\; g_{\alpha \alpha'}(x)\{H_m,\rho^{\mu \mu'}_W(t) c_{\mu \mu'}(x)\}_{X,x} \nonumber \\
&&\quad   - \frac{\hbar}{8} \int dx\; g_{\alpha \alpha'}(x)\left[\frac{\partial h^{\lambda\lambda'}}{\partial R}\Big(\frac{\partial^2}{\partial r_{\lambda} \partial r_{\lambda'}}+ \frac{\partial^2}{\partial p_{\lambda} \partial p_{\lambda'}}\Big)\cdot \frac{\partial }{\partial P}\right]\nonumber \\
&& \quad \times \rho^{\mu \mu'}_W(t) c_{\mu \mu'}(x).
\end{eqnarray}
In order to proceed with our analysis, we consider separately the Poisson bracket and second complex terms on the right hand side.

\subsection{Transformation of the Poisson bracket term}
The first contribution in the Poisson bracket, $\{H_m,\rho_m(t)\}_{X,x}$, arising from derivatives with respect to the mapping variables is
$- \frac{h^{\lambda\lambda'}}{\hbar}\left(p_{\lambda'} \frac{\partial}{\partial r_{\lambda}}-r_{\lambda'} \frac{\partial}{\partial p_{\lambda}} \right) \rho_m$ (see Eq.~(\ref{ultimate_expression_for_QCL1})). Its back transformation is given by
\begin{equation} \label{FT11-step1}
PB1= \int drdp \; g_{\alpha\alpha'} (- \frac{h^{\lambda\lambda'}}{\hbar}) \left( p_{\lambda'} \frac{\partial}{\partial r_{\lambda}}-r_{\lambda'} \frac{\partial}{\partial p_{\lambda}} \right) \rho^{\mu\mu'}_W c_{\mu\mu'}
\end{equation}
Direct substitution of the expressions for $g_{\alpha\alpha'}$ and $f_{\mu\mu'}$ results in a sum of integrals that can be evaluated analytically (they all are in the form of a polynomial times a Gaussian function) to obtain
\begin{equation}\label{FT11}
PB1= -i\tilde \omega_{\alpha\alpha'}\rho_W^{\alpha\alpha'} + \frac{i}{\hbar}\big( \rho_W^{\alpha\alpha''} V_c^{\alpha''\alpha}-V_c^{\alpha\alpha''}\rho_W^{\alpha''\alpha} \big)
\end{equation}
after substitution of ${h}^{\alpha\alpha'}$ from Eq.~(\ref{hamiltonian-matrix}). This contribution  represents the evolution of the quantum subsystem and includes the influence of the environmental degrees of freedom through the coupling potential $\hat{V}_c$.

Next, we consider the back transformation of the second contribution in Eq.~(\ref{ultimate_expression_for_QCL1}) coming from derivatives with respect to environmental coordinates,
$\big(\frac{\partial H_m}{\partial R}\cdot  \frac{\partial }{\partial P}
-\frac{P}{M}\cdot \frac{\partial }{\partial R} \big) \rho_m(t)$. Given the form of $H_m$ in Eq.~(\ref{h_m-def}), one can write this contribution as
\begin{eqnarray}\label{second-term}
\frac{\partial H_m}{\partial R}\cdot  \frac{\partial \rho_m}{\partial P} -\frac{P}{M}\cdot \frac{\partial \rho_m}{\partial R}  &=& \frac{\partial V_e}{\partial R} \cdot \frac{\partial \rho_m}{\partial P} -\frac{P}{M}\cdot \frac{\partial \rho_m}{\partial R} \nonumber \\
&+&\frac{\partial h_m}{\partial R}\cdot  \frac{\partial \rho_m}{\partial P}.
\end{eqnarray}
Because $\partial V_e/\partial R=-F_e(R)$ and $P/M$ are independent of the mapping variables, the back transformation of the first two terms in Eq.~(\ref{second-term}) is simple:
\begin{eqnarray} \label{FT20-step1}
PB2&=&\int drdp\; g_{\alpha\alpha'} \Big(\frac{\partial V_e}{\partial R}\cdot  \frac{\partial }{\partial P} -\frac{P}{M} \cdot \frac{\partial}{\partial R}\Big)\rho^{\lambda\lambda'}_W c_{\lambda\lambda'} \nonumber \\
&& \qquad = -\Big(F_e \cdot \frac{\partial}{\partial P} +\frac{P}{M} \cdot \frac{\partial }{\partial R}
\Big) \rho^{\alpha\alpha'}_W.
\end{eqnarray}
The last term in Eq.~(\ref{second-term}), arising from $\frac{\partial h_m}{\partial R}$, gives
\begin{eqnarray} \label{FT21}
PB3&=&\int drdp\; g_{\alpha\alpha'} \frac{1}{2\hbar} (r_{\lambda}r_{\lambda'} + p_{\lambda}p_{\lambda'}-\hbar\delta_{\lambda\lambda'}) \nonumber \\
&\times& \frac{\partial h^{\lambda \lambda'}}{\partial R} \cdot \frac{\partial \rho^{\mu\mu'}_W}{\partial P} c_{\mu\mu'}(x)
\end{eqnarray}
This integral too can be performed after substitution of the $g_{\alpha\alpha'}$ and $c_{\mu\mu'}$ functions to get
\begin{eqnarray}\label{FT20}
PB3&=&  \frac{1}{2}\Big(\frac{\partial V_c^{\alpha\alpha''}}{\partial R}\cdot \frac{\partial \rho^{\alpha''\alpha'}_W}{\partial P}
+ \frac{\partial \rho^{\alpha\alpha''}_W}{\partial P} \cdot \frac{\partial V^{\alpha''\alpha'}_c}{\partial R}\Big)\nonumber \\
&&\quad +\frac{1}{4}  {\rm Tr}'\Big(\frac{\partial \hat{V}_c}{\partial R} \cdot \frac{\partial \hat{\rho}_W}{\partial P}\Big) \delta_{\alpha\alpha'},
\end{eqnarray}
where ${\rm Tr}'$ indicates a trace over the quantum subsystem states.

Combining the results above, we obtain
\begin{eqnarray}\label{poisson-back}
&&\int dx\; g_{\alpha \alpha'}(x)\big\{H_m,\rho_m(t)\big\}_{X,x}= -i\tilde{\mathcal L}_{\alpha \alpha', \mu \mu'} \rho^{\mu \mu'}_W(t)\nonumber \\
&&\qquad \qquad +\frac{1}{4}  {\rm Tr}'\Big(\frac{\partial \hat{V}_c}{\partial R} \cdot \frac{\partial \hat{\rho}_W}{\partial P}\Big) \delta_{\alpha\alpha'} .
\end{eqnarray}
Thus, we find that the back transformation of the Poisson bracket term yields the QCL operator in the subsystem basis defined in Eq.~(\ref{QCLE in SSB}), plus an extra contribution $\frac{1}{4}  {\rm Tr}'\Big(\frac{\partial \hat{V}_c}{\partial R} \cdot \frac{\partial \hat{\rho}_W}{\partial P}\Big) \delta_{\alpha\alpha'}$. It is this term that is responsible for any errors in the approximate simulations of the QCLE that include only the Poisson bracket term.

An understanding of the physical meaning of this extra contribution can be obtained from the following considerations: First, we observe that the trace of the partially Wigner transformed density
 matrix over the quantum subsystem states is just the phase space density of the environment, $\rho_b(X,t)={\rm Tr}'\hat{\rho}_W(X,t)$. Taking the trace of Eq (\ref{QCLE in SSB}) yields,
\begin{eqnarray}
\frac{\partial }{\partial t}\rho_b(X,t) &=& - \Big( F_e(R)\cdot \frac{\partial}{\partial P}+\frac{P}{M} \cdot \frac{\partial}{\partial R}\Big){\rho}_b(t)\nonumber \\
&&+ {\rm Tr}'\Big(\frac{\partial \hat{V}_c}{\partial R}\cdot  \frac{\partial \hat{\rho}_W(t)}{\partial P}\Big),
\end{eqnarray}
or, equivalently,
\begin{eqnarray}\label{eq:trace}
{\rm Tr}'\Big(\frac{\partial \hat{V}_c}{\partial R}\cdot  \frac{\partial \hat{\rho}_W(t)}{\partial P}\Big)&=& \Big(\frac{\partial }{\partial t}+ F_e(R)\cdot \frac{\partial}{\partial P}+\frac{P}{M}\cdot  \frac{\partial}{\partial R}\Big)
{\rho}_b(t)\nonumber \\
&\equiv&\Big(\frac{\partial }{\partial t}+ i L_e\Big){\rho}_b(t),
\end{eqnarray}
where $i L_e$ is the classical Liouville operator for the environment in isolation from the quantum subsystem.
Consequently, the trace term in Eq.~(\ref{eq:trace}) can be interpreted as the time variation of the phase space density of the environment along the flow lines generated by the classical evolution of the environment in isolation from the quantum subsystem. In a system where there is no coupling between the environment and the quantum subsystem, this term is zero by Liouville's theorem. As a result, a non-zero value of the trace contribution can be interpreted as an effect arising from the back reaction of the quantum subsystem on its environment.

\subsection{Excess coupling term}
To complete the back transformation of the mapping QCLE to the subsystem basis, we must back transform the  complex last term in Eq.~(\ref{map-QCL1}):
\begin{eqnarray}
\label{FT40}
&&-\frac{\hbar}{8}\int drdp \; g_{\alpha\alpha'}  \left[\frac{\partial h^{\lambda\lambda'}}{\partial R}\big(\frac{\partial^2}{\partial r_{\lambda} \partial r_{\lambda'}}+ \frac{\partial^2}{\partial p_{\lambda} \partial p_{\lambda'}}\big)\cdot \frac{\partial }{\partial P}\right]\nonumber \\
&&\quad \times  \rho^{\mu\mu'}_W c_{\mu\mu'} \nonumber \\
&&= -\frac{1}{8} \Big(\frac{\partial h^{\lambda\lambda'}}{\partial R}\cdot  \frac{\partial \rho^{\lambda\lambda'}_W}{\partial P} + \frac{\partial h^{\lambda\lambda'}}{\partial R} \cdot \frac{\partial \rho^{\lambda'\lambda}_W}{\partial P} \Big)\delta_{\alpha\alpha'} \nonumber \\
&&= -\frac{1}{4} {\rm Tr}'\big(\frac{\partial \hat{V}_c}{\partial R}\cdot  \frac{\partial \hat{\rho}_W}{\partial P}\big)\delta_{\alpha\alpha'}.
\end{eqnarray}
Since this result is again proportional to ${\rm Tr}'\Big(\frac{\partial \hat{V}_c}{\partial R} \cdot \frac{\partial \hat{\rho}_W(t)}{\partial P}\Big)$, we call this contribution a excess coupling term.

Inserting these results into Eq.~(\ref{map-QCL1-back}), the QCLE in the subsystem basis (Eq.~(\ref{QCLE in SSB})) is obtained as expected. However, from this derivation, we learn that the excess coupling term is equal and opposite in sign to the last term in Eq.~(\ref{poisson-back}). As a result, this contribution exactly cancels the analogous term in the back-transformed Poisson bracket expression to yield the desired result.

\section{Approximate mapping QCLE \label{sec:approx}}

The results of the previous section can be used to assess the utility of the approximate mapping QCLE where only the Poisson bracket contribution is retained in the Liouville operator. This Poisson bracket mapping equation (PBME) is
\begin{eqnarray}
\label{ultimate_expression_for_QCL2}
\frac{\partial}{\partial t} \rho_m(t) & \approx &\{H_m,\rho_m(t)\}_{X,x}\nonumber \\
&=& - \frac{h^{\lambda\lambda'}}{\hbar} \left(p_{\lambda'} \frac{\partial}{\partial r_{\lambda}}-r_{\lambda'} \frac{\partial}{\partial p_{\lambda}} \right) \rho_m(t) \nonumber\\
&&+\frac{\partial H_m}{\partial R}\cdot  \frac{\partial \rho_m(t)}{\partial P}-\frac{P}{M}\cdot \frac{\partial \rho_m(t)}{\partial R}.
\end{eqnarray}
It can be solved by characteristics~\cite{hyo:ray:ali} and the resulting set of ordinary differential equations is~\cite{footnote1}
\begin{eqnarray} \label{solution}
\frac{d r_\lambda(t)}{dt} &=& \frac{1}{\hbar}h^{\lambda\lambda'}(R(t))p_{\lambda'}(t),\nonumber \\
\frac{d p_\lambda(t)}{dt} &=& -\frac{1}{\hbar}h^{\lambda\lambda'}(R(t))r_{\lambda'}(t), \nonumber \\
\frac{d R(t)}{dt} &=& \frac{P(t)}{M},  \hspace{1.0cm} \frac{dP(t)}{dt} = - \frac{\partial H_m}{\partial R(t)}.
\end{eqnarray}
Consequently, the simulation of the dynamics described by Eq.~(\ref{ultimate_expression_for_QCL2}) is an easy task.~\cite{kelly10,footnote2}

From the results in Sec.~\ref{transformation} it follows that Eq.~(\ref{ultimate_expression_for_QCL2}) is equivalent to the following equation in the subsystem basis:
\begin{eqnarray}\label{QCL-poisson-approx}
\frac{\partial}{\partial t} \rho_W^{\alpha \alpha'}(X,t)&\approx& -i\tilde{\mathcal L}_{\alpha \alpha', \mu \mu'} \rho^{\mu \mu'}_W(t) \nonumber \\
&&+\frac{1}{4}  {\rm Tr}'\Big(\frac{\partial \hat{V}_c}{\partial R} \cdot \frac{\partial \hat{\rho}_W(t)}{\partial P}\Big) \delta_{\alpha\alpha'}.
\end{eqnarray}
Because of the presence of the second term on the right side, this form of the equation shows that the back reaction of the quantum subsystem on the dynamics of the environmental degrees of freedom is incorrectly described and the solution by characteristics is only an approximation to full QCL dynamics. Two features should be kept in mind when considering the error incurred in the use of the PBME. First, the effects of the environment on the quantum subsystem are fully accounted for in the Poisson bracket, and the effects of the quantum subsystem on the environment are also included in this term through the first term in PB3 in Eq.~(\ref{FT20}). Second, note the factor of $1/4$ in Eq.~(\ref{QCL-poisson-approx}): the error involves only a portion of the back reaction of the quantum subsystem on the evolution of the environmental density. To the extent that the effects of the excess coupling term are small, simulations employing the PBME for the evolution will provide an accurate description of the dynamics.

The equivalence of Eqs.~(\ref{QCL-poisson-approx}) and (\ref{ultimate_expression_for_QCL2}) suggests a means to gauge the importance of the excess coupling term. Rather than focussing on the excess coupling term itself, it is more convenient to consider its effect on the estimates of expectation values of observables. The expectation value of an arbitrary operator $\hat{B}_W(X)$ can be written as~\cite{hyo:ray:ali}
\begin{eqnarray}\label{eq:exvalue}
\overline{B(t)}&=& \int dX\; B_{W}^{\lambda\lambda'}(X)\rho_{W}^{\lambda'\lambda}(X,t)\nonumber \\
&=& \int dX\; B_{W}^{\lambda\lambda'}(X,t)\rho_{W}^{\lambda'\lambda}(X)\nonumber \\
 & = & \int dXdx\; B_{m}(x,X,t)\tilde{\rho}_{m}(x,X),\label{barB-map}
\end{eqnarray}
In the first line of Eq.~(\ref{eq:exvalue}) the expectation value is computed in the subsystem basis, in the second line the time evolution is moved from the density matrix to the operator, while the last line gives the expectation value computed in the mapping basis. The quantities entering in the mapping form of the expectation value are the time dependent observable,
\begin{eqnarray}\label{eq:Beq}
&& B_m(x,X,t)=(2 \hbar)^{-1}B^{\lambda \lambda'}_W(X(t))\Big(r_\lambda(t) r_{\lambda'}(t) \\
&& \quad +p_\lambda(t) p_{\lambda'}(t)  + i(r_\lambda(t) p_{\lambda'}(t)-p_\lambda(t) r_{\lambda'}(t))-\hbar \delta_{\lambda \lambda'}\Big),\nonumber
\end{eqnarray}
and the initial density $\tilde{\rho}_{m}(x,X)=\int dx'f(x,x')\rho_{m}(x',X)$ with
\begin{eqnarray}\label{eq:fxxp}
 &  & f(x,x')=\frac{1}{(2\pi\hbar)^{N}}\int dzdz'\langle m_{\lambda}|r-\frac{z}{2}\rangle\langle r+\frac{z}{2}|m_{\lambda'}\rangle\nonumber \\
 &  & \qquad\times\langle m_{\lambda'}|r'-\frac{z'}{2}\rangle\langle r'+\frac{z'}{2}|m_{\lambda}\rangle e^{-i(p\cdot z+p'\cdot z')/\hbar}.
\end{eqnarray}
The integrals involved in the definition of the function ${f(x,x')}$ in Eq.~(\ref{eq:fxxp}) can be performed analytically. For a two-level system the result is
\begin{eqnarray} \label{f(x,x')2}
f(x,x')&=& \frac{8}{\pi^2\hbar^4}\Big[2(r\cdot r' +p \cdot p')^2 +2(r \cdot p' - r' \cdot p)^2 \nonumber \\
&&  -\hbar (x^2 + x^{\prime 2}) +\hbar^2 \Big] e^{-(x^2+x^{\prime 2})/\hbar}, \nonumber
\end{eqnarray}
where $r \cdot r' =r_1 r_1' +r_2 r_2'$, with similar expressions for other scalar products. Thus, these formulas allow one to use either the mapping or subsystem bases to compute the expectation values of any observable.

While Eq.~(\ref{eq:exvalue}), along with Eqs.~(\ref{eq:Beq}) and (\ref{solution}), provide a simple means to compute the expectation value of an operator using the PBME, it is convenient to consider the computation of the general operator $\hat{B}_W(X)$ using the PBME written in the subsystem basis in order to gain further insight into the effects of the excess coupling term on such average values. Taking the time derivative of Eq.~(\ref{eq:exvalue}) and using Eq.~(\ref{QCL-poisson-approx}) we find
\begin{eqnarray}\label{Bav-PBME}
\frac{d  \overline{B(t)}}{d t} &\approx& -i\int dX \; B_{W}^{\alpha' \alpha}(X)\tilde{\mathcal L}_{\alpha \alpha', \mu \mu'} \rho^{\mu \mu'}_W(t) \\
&& \quad +\frac{1}{4}  \int dX\; B_{W}^{\alpha' \alpha}(X){\rm Tr}'\Big(\frac{\partial \hat{V}_c}{\partial R}\cdot  \frac{\partial \hat{\rho}_W(t)}{\partial P}\Big) \delta_{\alpha\alpha'}, \nonumber \\
&=& -i\int dX \; B_{W}^{\alpha' \alpha}(X)\tilde{\mathcal L}_{\alpha \alpha', \mu \mu'} \rho^{\mu \mu'}_W(t) \nonumber \\
&& \quad -\frac{1}{4}  \int dX\; \frac{\partial B_{W}^{\alpha' \alpha}(X)}{\partial P}\cdot \frac{\partial V^{\lambda \lambda'}_c}{\partial R}  \rho^{\lambda' \lambda}_W(t) \delta_{\alpha\alpha'}, \nonumber
\end{eqnarray}
where, as usual, the Einstein summation convention is used. In the last line of this equation an integration by parts with respect to $P$ was performed so that the momentum derivative of the density matrix no longer appears. Therefore, the effect of the excess coupling term on the evolution of the expectation value of $\hat{B}_W(X)$ can be estimated by computing the average values on the right side of this equation and determining their relative magnitudes.

If the initial value of the operator $\hat{B}_W(X)$ is a function only of the configuration space coordinates, or independent of environmental phase space coordinates, then the last excess coupling term in Eq.~(\ref{Bav-PBME}) will vanish because the momentum derivative is zero, and for this case we have
\begin{eqnarray}\label{popav-PBME}
\frac{d  \overline{B(t)}}{d t} = -i\int dX \; B_{W}^{\alpha' \alpha}(X)\tilde{\mathcal L}_{\alpha \alpha', \mu \mu'} \rho^{\mu \mu'}_W(t).
\end{eqnarray}
For such observables the excess coupling term will enter the computation of the average value indirectly through the density matrix on the right side. This can be seen by writing the formal solution of Eq.~(\ref{QCL-poisson-approx}) as
\begin{eqnarray}
&&\rho_W^{\alpha \alpha'}(t)= \Big(e^{-i \tilde{{\mathcal L}}t} \Big)_{\alpha \alpha', \nu \nu'}\rho_W^{\nu \nu'}(0) \\
&&\qquad +\int_0^t dt' \; \Big(e^{-i \tilde{{\mathcal L}}(t-t')} \Big)_{\alpha \alpha', \nu \nu'} \frac{1}{4} \frac{\partial V_c^{\mu \mu'}}{\partial R} \cdot \frac{\partial \rho_W^{\mu' \mu}(t)}{\partial P} \delta_{\nu \nu'},\nonumber
\end{eqnarray}
and inserting the result on the right of Eq.~(\ref{popav-PBME}). The evaluation of this higher order correction is difficult because of the convolution in the above equation. In particular these considerations apply to the computation of important observables such as the quantum subsystem populations and quantum coherence, since the initial values of these observables are independent of the phase space coordinates.

If instead the observable is a function $G(P)$ only of the environmental momenta, then the excess coupling term enters the computation of the time rate of change of its expectation value directly.  Consider the expectation value of $B_{W}^{\alpha' \alpha}(X)=G(P) \delta_{\alpha \alpha'}$,
\begin{eqnarray}
\overline{G(t)}=\sum_\alpha \int dX\; G(P) \rho^{\alpha \alpha}_W(X,t).
\end{eqnarray}
If we compute the time derivative of this quantity using Eq.~(\ref{Bav-PBME}) we obtain,
\begin{eqnarray}\label{eq:G-evol}
\frac{d \overline{G(t)}}{dt}&=& \int dX\; \frac{\partial G(P)}{\partial P} \cdot F^{\alpha \alpha'} \rho^{\alpha' \alpha}_W(X,t)  \\
&& +\frac{N }{4} \int dX\; \frac{\partial G(P)}{\partial P} \cdot F^{\alpha \alpha'}_c \rho^{\alpha' \alpha}_W(X,t),\nonumber
\end{eqnarray}
where $F^{\alpha \alpha'}=-\partial (V_e(R)+V_c^{\alpha \alpha'}(R))/\partial R$ is the total force on the environment and $F_c^{\alpha \alpha'}=-\partial V_c^{\alpha \alpha'}(R)/\partial R$ is the force contribution arising from the quantum subsystem-environment coupling. The first contribution on the right side of Eq.~(\ref{eq:G-evol}) is obtained from the explicit computation of $-i\int dX \; G(P) \delta_{\alpha' \alpha}\tilde{\mathcal L}_{\alpha \alpha', \mu \mu'} \rho^{\mu \mu'}_W(t)$, while the second contribution comes from the evaluation of $\frac{1}{4}  \int dX\; G(P) \delta_{\alpha' \alpha}{\rm Tr}'\Big(\frac{\partial \hat{V}_c}{\partial R} \cdot \frac{\partial \hat{\rho}_W(t)}{\partial P}\Big) \delta_{\alpha\alpha'}$.
These formulas provide an indication of how the time evolution of the environmental momenta are influenced by the excess coupling term. In particular, if the coupling is weak or if the environment is large and only a portion of the environment is directly coupled to the quantum subsystem, the total force on the environment will dominate the coupling force and this effect will be small.

\section{Curve Crossing Models} \label{Numerical investigations}

In this section we consider the simulation of quantum-classical Liouville dynamics in the mapping basis for two curve crossing models studied earlier by Tully~\cite{Tully}. In particular we carry out simulations using the PBME and compare the results with numerically exact quantum dynamics  using a discrete Fourier transformation method~\cite{koslof}for these systems. Previous studies of the spin-boson model showed that the PBME yields results which are in excellent agreement with the numerically exact quantum simulations for the entire range of system parameters that were studied~\cite{hyo:ray:ali}. In the spin-boson model, the coupling between the environment and the quantum subsystem is linear in the environmental coordinates and is non-zero for all values of this quantity. The curve crossing models studied here provide a further test of the PBME for a situation where the coupling
 among the different degrees of freedom is nonlinear in the environmental coordinates and is localized in the configuration space of the environment.

\subsubsection*{A: Simple avoided crossing}
\begin{figure}[htbp]
\includegraphics[width=8.0cm, height=5cm]{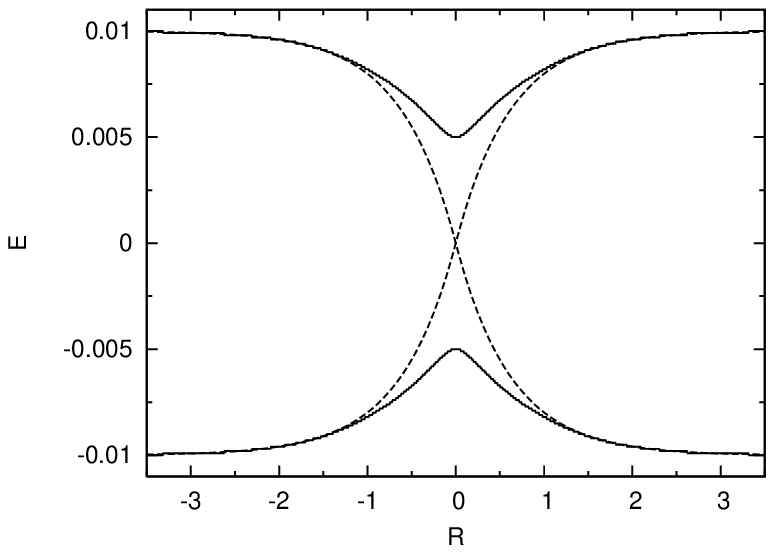}
\includegraphics[width=8.0cm, height=6.3cm]{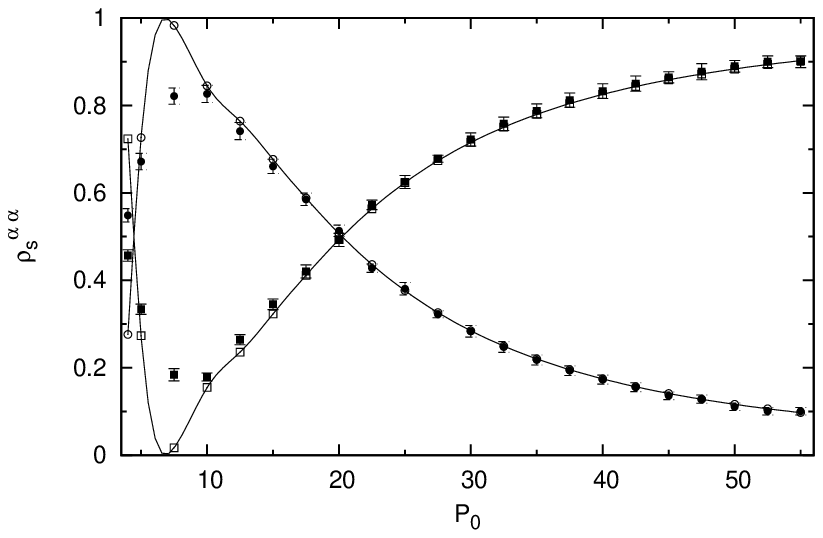}
\caption{(top) Diagonal elements of the Hamiltonian in the diabatic basis (dashed lines) and adiabatic (solid lines) energies for the simple avoided crossing model. (bottom) Asymptotic populations of state 1 (solid squares with error bars) and state 2 (solid circles with error bars) as a function of the initial momentum of the wave packet, $P_0$, for the simple avoided crossing. The corresponding numerically exact quantum results are indicated by open squares for state 1 and open circles for state 2.  The exact results are connected with lines as a guide for the eye.}\label{fig:simple}
\end{figure}

The simple two-state avoided crossing is defined by the following Hamiltonian matrix in the diabatic basis~\cite{Tully}:
\begin{equation}
{\bf h}=\left(\begin{array}{cc} \label{hamiltonian4}
A[1-e^{-B|R|}]\frac{R}{|R|} & Ce^{-DR^2} \\
Ce^{-DR^2} & -A[1-e^{-B|R|}]\frac{R}{|R|}\end{array}\right).
\end{equation}
The corresponding total Hamiltonian in the mapping basis is
\begin{eqnarray}
H_{m} &=& \frac{P^2}{2M} +\frac{1}{2\hbar} A\big[1-e^{-B|R|}\big] \frac{R}{|R|} (r_{1}^{2}+p_{1}^{2}-r_{2}^{2}-p_{2}^{2}) \nonumber \\
&&+ \frac{C}{\hbar}e^{-DR^2}(r_{1}r_{2}+p_{1}p_{2}).
\end{eqnarray}
The diagonal elements of the Hamiltonian in the diabatic basis and the adiabatic energies for this model are sketched in Fig.~\ref{fig:simple}. Our simulations are carried out using atomic units (a.u.) and the parameters in these units are taken to be $A=0.01$, $B=1.6$, $C=0.005$ and $D=1.0$.

We assume that the initial density matrix is uncorrelated and can be factored in to a product of subsystem and environmental contributions, i.e., $\tilde{\rho}_m(X,x)= \tilde{\rho_s}(x)\rho_e(X)$, where $\tilde{\rho}_s(x)$ is the subsystem density matrix in the mapping basis and $\rho_e(X)$ is the distribution function for the environment. The  density $\rho_e(X)$ is chosen to be the Wigner transform of a Gaussian wave packet centered at $R_0$ with momentum $P_0$:
\begin{equation}
\rho_e(X)=\frac{\sigma}{\sqrt{\pi}\hbar} e^{-(P-P_0)^2 \sigma^2/\hbar^2} \frac{1}{\sqrt{\pi}\sigma}e^{-(R-R_0)^2/\sigma^2}.
\end{equation}
The mass $M$ of the environmental degree of freedom is taken to be 2000 atomic mass units, while $R_0=-3.8$ a.u. and $\sigma=1$.

We assume the subsystem is initially in state 1, so that in the mapping basis
\begin{equation}
\tilde{\rho}_{s}(x) = \frac{2}{\pi^2\hbar^3} (r_{1}^{2}+p_{1}^{2}-\frac{\hbar}{2})e^{-x^2/\hbar}.
\end{equation}
We consider the time evolution of the subsystem populations, $\rho_s^{\alpha \alpha}(t),\; \alpha=1,2$, and coherence as measured by the off-diagonal element of the subsystem density matrix, $\rho_s^{12}(t)$. These quantities are conveniently computed using the formulas in Eq.~(\ref{eq:exvalue}) for a general operator by selecting $B_W^{\lambda \lambda'}= \delta_{\lambda \alpha} \delta_{\lambda' \alpha}$ and $B_W^{\lambda \lambda'}= \delta_{\lambda 1} \delta_{\lambda' 2}$, respectively. The asymptotic values of the populations in states 1 and 2 after the system has passed through the interaction region are shown in Fig.~\ref{fig:simple} as a function of the initial momentum of the wave packet. We see that the PBME and exact quantum results are in excellent agreement for values of $P_0$ above 10. Under this threshold, nuclear quantum effects play an important role so the most likely cause for the discrepancies in the figure is the breakdown of the classical dynamics approximation for the nuclei (note that usually, mixed quantum-classical non-adiabatic tests do not explore the range below this value of the momentum for this model).

\begin{figure}[htbp]
\includegraphics[width=8.0cm]{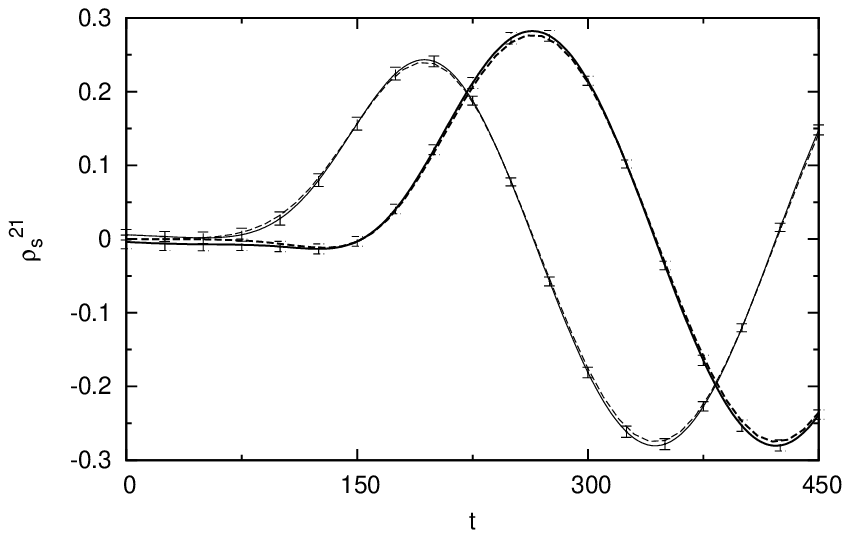}
\includegraphics[width=8.0cm]{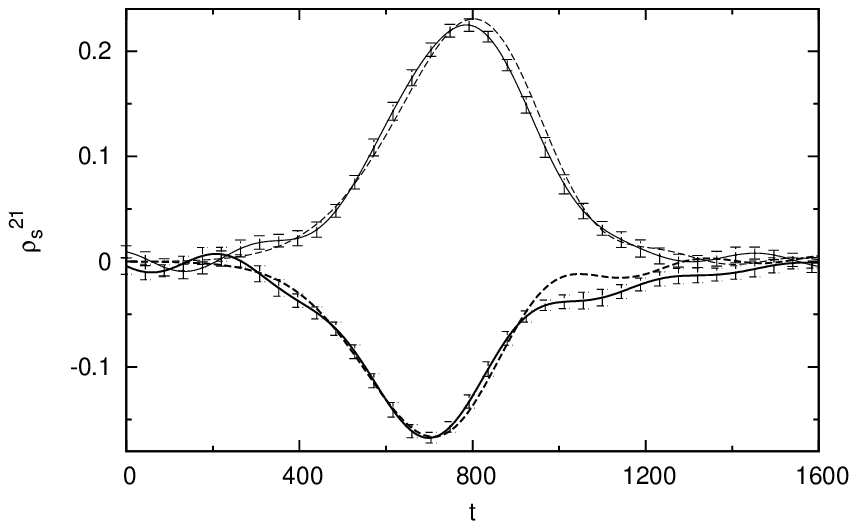}
\caption{The real (thick lines) and imaginary (thin lines) parts of $\rho_s^{12}(t)$ versus time for the simple avoided crossing model. The dashed lines denote the full quantum results while the solid lines are the results of computations using the PBME. (top) $P_0=50$, (bottom) $P_0=10$.}
\label{fig:coherence-simp}
\end{figure}

The time evolution of the real and imaginary parts of $\rho_s^{12}(t)$ are displayed in Fig.~\ref{fig:coherence-simp}. 
These results for the quantum coherence are also in good agreement with the numerically exact full quantum simulations.
The comparisons shown in these figures test two effects at the same time: the validity of the QCLE and its approximation by the PBME. Recall that from the considerations in Sec.~\ref{sec:approx} the effects of the excess coupling term enter the populations and quantum coherence only as higher order contributions.

For the simple avoided crossing (and the dual avoided crossing considered below), $V_e(R)=0$ so that $F^{\alpha \alpha'}=F_c^{\alpha \alpha'}$ and the two integrals on the right of Eq.~(\ref{eq:G-evol}) are equal, and the two contributions differ only in their prefactors. For this two-level model $N=2$ and the PBME prediction for the time rate of change of $\overline{G(t)}$ is $3/2$ that of QCL dynamics. This is a severe test of the approximation since $V_e(R)=0$ for this model. Thus, there is a substantial influence on the environmental momenta but, as noted above, these enter only in higher order corrections to the quantum subsystem populations and coherences, contributing to the good agreement with the exact results seen in the figures.

\subsubsection*{B: Dual avoided crossing}

\begin{figure}
\includegraphics[width=8.0cm, height=5cm]{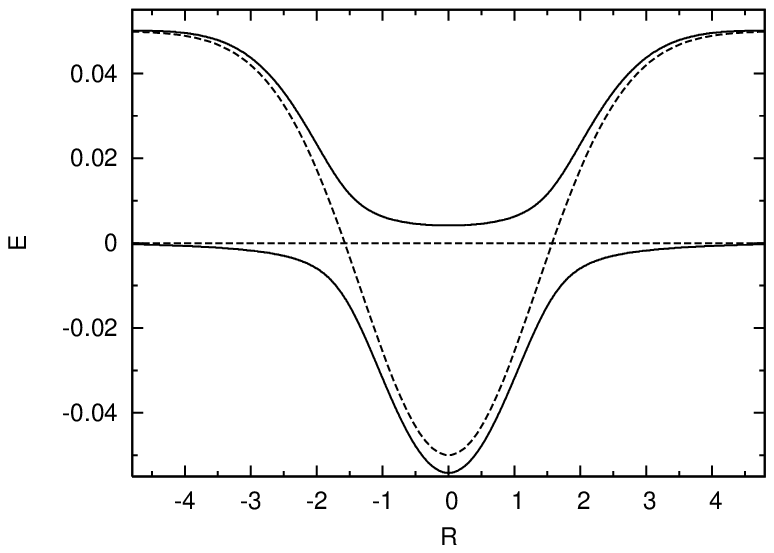}
\includegraphics[width=8.0cm, height=6.3cm]{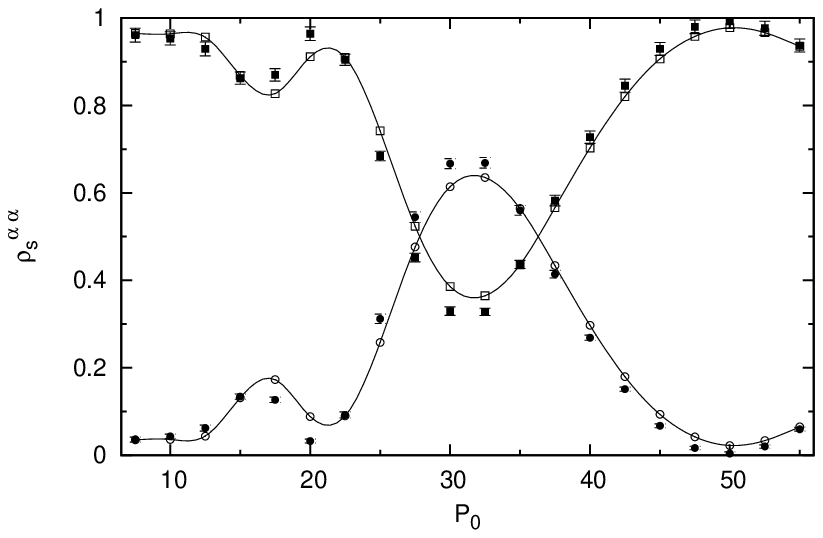}
\caption{(top) Diabatic diagonal elements of the Hamiltonian (dashed lines) and adiabatic energies (solid lines) for the dual avoided crossing model. (bottom) Populations of the diabatic states with same symbols as those in Fig.~\ref{fig:simple}.}\label{fig:dual}
\end{figure}

The dual avoided crossing model provides a further test of the theory. The Hamiltonian matrix in the diabatic basis takes the form~\cite{Tully},
\begin{equation} \label{hamiltonian5}
{\rm{\bf  h}}=\left(\begin{array}{cc}
0 & Ce^{-DR^2}\\
Ce^{-DR^2} & -Ae^{-BR^2} + E_0\end{array}\right),
\end{equation}
resulting in the mapping Hamiltonian,
\begin{eqnarray}
H_{m} &=& \frac{P^2}{2M} +\frac{1}{2\hbar} (-Ae^{-BR^2} + E_0)(r_{2}^{2}+p_{2}^{2}-1) \nonumber \\
&&+ \frac{C}{\hbar}e^{-DR^2}(r_{1}r_{2}+p_{1}p_{2}).
\end{eqnarray}
The system parameters were taken to be $A=0.10$, $B=0.28$, $C=0.015$, $E_0=0.05$ and $D=0.06$, again in atomic units. The initial conditions have the same form as for the simple avoided crossing, with $R_0=-10$.
The diagonal matrix elements in the diabatic representation and the adiabatic energies for this model are plotted in Fig.~\ref{fig:dual} (top) and the populations as function of $P_0$ in the bottom panel of this figure.
One can see that the agreement with exact quantum dynamics is very good over the entire range of $P_0$, although there are small discrepancies in the magnitudes of maxima and minima in the population curves.

\begin{figure}[htbp]
\includegraphics[width=8.0cm]{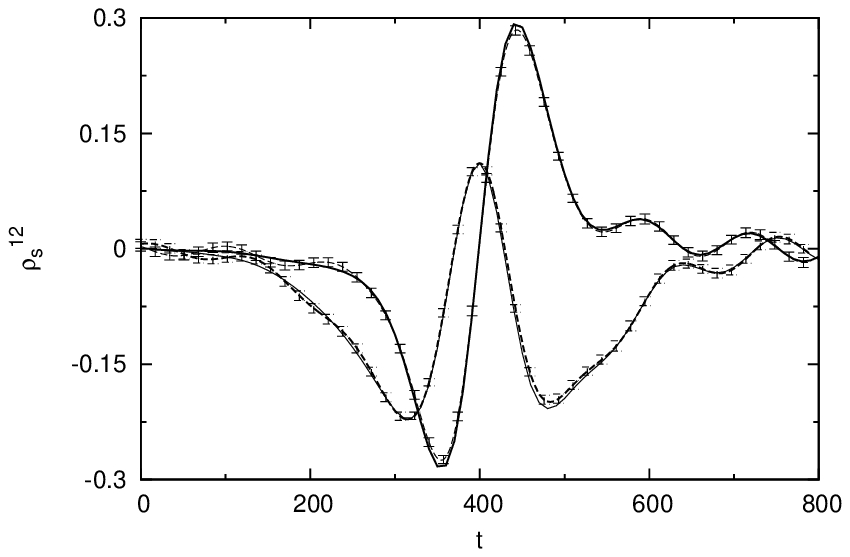}
\includegraphics[width=8.0cm]{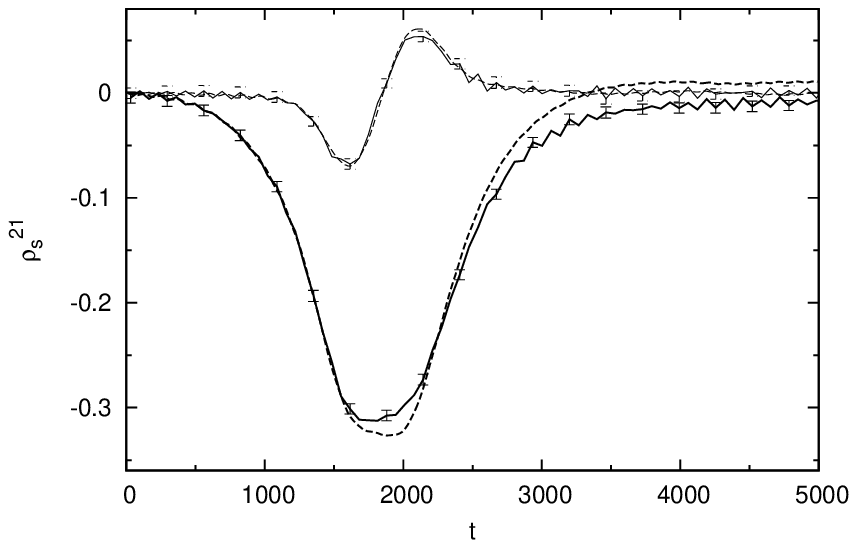}
\caption{The real and imaginary parts of $\rho_s^{12}(t)$ versus time for the dual avoided crossing model. Same line types as in Fig.~\ref{fig:coherence-simp}. (top) $P_0=50$, (bottom) $P_0=10$.}
\label{fig:coherence-dual}
\end{figure}

The time evolution of the real and imaginary parts of $\rho_s^{12}(t)$ for the dual avoided crossing model are shown in Fig.~\ref{fig:coherence-dual}.
The results for the quantum coherence are again in good agreement with the numerically exact full quantum simulations for this model.

\section{Conclusion} \label{Conclusion}

By transforming the mapping form of the QCLE term-by-term back to the subsystem basis we were able to identify the nature of the complex term in this equation that makes its simulation difficult. This observation then led to an analysis of the PBME, the approximate form of the mapping QCLE that retains only the Poisson bracket term in the Liouville operator. The PBME is easily simulated by classical trajectories. However, when compared to the full QCLE, it contains an extra term that can be interpreted as being proportional to the effect of the quantum subsystem on the time evolution of the density matrix of the environment along the flow lines generated by the classical evolution of the environment in isolation from the quantum subsystem. The excess coupling term in the full mapping QCLE cancels this contribution to yield the original QCLE.

Earlier simulations~\cite{stotho,ThoSto99,hyo:ray:ali} on spin-boson model with bilinear coupling to the bath, along with the present simulations on two curve crossing models with nonlinear bath coupling, indicate that the correction terms are small and the approximate PBME yields good agreement with the exact quantum results for these models. These results suggest that this approximate evolution equation will be useful in many applications; however, the validity of the PBME must be tested for the specific problem under consideration and it may not always be a good approximation to full QCL dynamics~\cite{footnote3}. The effects of the excess coupling term on the time evolution of the observables can be estimated by computing the expectation values in Eq.~(\ref{Bav-PBME}) and these values  provide one way of gauging the importance of the deviations from the QCLE. Although care must be exercised in the use of the PBME when applied to a given system, the ease with which it can be simulated and its accuracy for many systems of interest make it a powerful simulation scheme that should find increased use in future applications.

\subsubsection*{Acknowledgement}
This work was supported in part by a grant from the Natural Sciences and Engineering Council of Canada. SB acknowledges funding from a SEED grant from IIT-Genova.


\section*{Appendix}

In this Appendix we show how $c_{\lambda\lambda'}(x)$ in Eq.~(\ref{FD}) and $g_{\lambda\lambda'}(x)$ in Eq.~(\ref{GD}) can be computed.

\subsection*{Calculation of $c_{\lambda\lambda'}(x)$}
Starting from its definition we have,
\begin{widetext}
\begin{eqnarray} \label{AFD}
&&(2\pi\hbar)^N c_{\lambda\lambda'}(x) = (\hat a_{\lambda}^\dag \hat a_{\lambda'})_W = \frac{1}{2\hbar} \int dz e^{ip \cdot z/\hbar} \langle r-\frac{z}{2}|[\hat{q}_{\lambda}\hat{q}_{\lambda'}  + \hat{p}_{\lambda}\hat{p}_{\lambda'} + i(\hat{p}_{\lambda}\hat{q}_{\lambda'}-\hat{q}_{\lambda}\hat{p}_{\lambda'})]|r+\frac{z}{2}\rangle \nonumber \\
&&
= \frac{1}{2\hbar} \int dz e^{ip \cdot z/\hbar} \Big[ (r-\frac{z}{2})_\lambda (r+\frac{z}{2})_{\lambda'} - \hbar^2 \frac{\partial}{\partial z_{\lambda}}\frac{\partial}{\partial z_{\lambda'}}  + \hbar(r+\frac{z}{2})_{\lambda'}\frac{\partial}{\partial z_{\lambda}} - \hbar(r-\frac{z}{2})_{\lambda}\frac{\partial}{\partial z_{\lambda'}} \Big] \delta(z)\nonumber \\
&& = \frac{1}{2\hbar} \Big[ r_\lambda r_{\lambda'} - \hbar^2 \int dz \delta(z) \frac{\partial}{\partial z_{\lambda}}\frac{\partial}{\partial z_{\lambda'}} e^{ip \cdot z/\hbar}  - \hbar\int dz \delta(z) \frac{\partial}{\partial z_{\lambda}} \big[(r+\frac{z}{2})_{\lambda'} e^{ip \cdot z/\hbar} \big] + \hbar\int dz \delta(z) \frac{\partial}{\partial z_{\lambda'}} \big[(r-\frac{z}{2})_{\lambda}e^{ip \cdot z/\hbar}\big] \Big] \nonumber \\
&& = \frac{1}{2\hbar} \Big[r_\lambda r_{\lambda'} - \hbar^2 (\frac{i}{\hbar}p_{\lambda}) (\frac{i}{\hbar}p_{\lambda'}) - \hbar(\frac{\delta_{\lambda\lambda'}}{2}+\frac{i}{\hbar}p_{\lambda}r_{\lambda'}) + \hbar(-\frac{\delta_{\lambda\lambda'}}{2}+\frac{i}{\hbar}p_{\lambda'}r_{\lambda})\Big] \nonumber \\
&& = \frac{1}{2\hbar} \Big[r_\lambda r_{\lambda'} + i(r_\lambda p_{\lambda'} - r_{\lambda'} p_{\lambda}) + p_{\lambda}p_{\lambda'} - \hbar\delta_{\lambda\lambda'} \Big].
\end{eqnarray}

\subsection*{Calculation of $g_{\lambda\lambda'}(x)$}
Similarly for $g_{\lambda \lambda'}(x)$ we have

\begin{eqnarray} \label{AGD}
&& g_{\lambda \lambda'}(x) = \int dz e^{-ip \cdot z/\hbar} \langle m_{\lambda}|r-\frac{z}{2}\rangle \langle r+\frac{z}{2}|m_{\lambda'}\rangle \nonumber \\
&& = \int dz e^{-ip \cdot z/\hbar} \phi_0(r_1-\frac{z_1}{2})\cdots \phi_1(r_\lambda-\frac{z_\lambda}{2})\cdots \phi_0(r_N-\frac{z_N}{2}) \phi_0(r_1+\frac{z_1}{2})\cdots \phi_1(r_{\lambda'}+\frac{z_{\lambda'}}{2})\cdots \phi_0(r_N+\frac{z_N}{2}) \nonumber \\
&& = \Big(\frac{1}{\pi\hbar}\Big)^{N/2} \frac{2}{\hbar} \int dz e^{-ip \cdot z/\hbar} \bigg\{ \delta_{\lambda\lambda'} e^{[- \frac{1}{2\hbar}((r_1-\frac{z_1}{2})^2+(r_1+\frac{z_1}{2})^2)]} \cdots (r_\lambda^2-\frac{z_\lambda^2}{4})  e^{[- \frac{1}{2\hbar}((r_\lambda -\frac{z_\lambda}{2})^2+(r_\lambda+\frac{z_\lambda}{2})^2)]} \nonumber \\
&&\qquad \times \cdots e^{[- \frac{1}{2\hbar}((r_N-\frac{z_N}{2})^2+(r_N+\frac{z_N}{2})^2)]} + (1-\delta_{\lambda\lambda'}) e^{[- \frac{1}{\hbar}\big(r_1^2+\frac{z_1^2}{4})]} \cdots (r_\lambda-\frac{z_\lambda}{2}) e^{[- \frac{1}{\hbar}\big(r_\lambda^2 + \frac{z_\lambda^2}{4})]} \cdots (r_{\lambda'}+\frac{z_{\lambda'}}{2}) \nonumber \\
&& \qquad \times e^{[- \frac{1}{\hbar}(r_{\lambda'}^2 + \frac{z_{\lambda'}^2}{4})]} \cdots e^{[- \frac{1}{\hbar}(r_N^2+\frac{z_N^2}{4})]} \bigg\} \nonumber\\
&& = \frac{2^{N+1}}{\hbar} e^{-x^2/\hbar} \Big[ \delta_{\lambda\lambda'} (r_\lambda^2-\frac{1}{4}(2\hbar-4p_\lambda^2))  + (1-\delta_{\lambda\lambda'}) (r_\lambda +ip_{\lambda}) (r_{\lambda'}-ip_{\lambda'}) \Big] \\
&& = \frac{2^{N+1}}{\hbar} e^{-x^2/\hbar}  \Big[r_\lambda r_{\lambda'} - i(r_\lambda p_{\lambda'} - r_{\lambda'} p_{\lambda}) + p_{\lambda}p_{\lambda'} -\delta_{\lambda\lambda'} \frac{\hbar}{2} \Big]  \nonumber,
\end{eqnarray}

where we have used Eq.~(\ref{map}) together with $\phi_0(x)= (\frac{1}{\pi\hbar})^{1/4}e^{-\frac{1}{2\hbar}x^2}, \hspace*{0.2cm} \phi_1(x) = (\frac{1}{\pi\hbar})^{1/4}\sqrt{\frac{2}{\hbar}} x e^{-\frac{1}{2\hbar}x^2}$. In this expression recall that $z=(z_1,z_2, \dots, z_N)$, $p=(p_1,p_2, \dots, p_N)$ and $p \cdot z=p_1z_1+p_2z_2+ \dots +p_Nz_N$, with similar vector notation for other unscripted quantities.
\end{widetext}


\begin{thebibliography}{49}
\bibitem[{foo({\natexlab{a}})}]{footnote-units}
\bibinfo{note}{Since the mapping oscillators are fictitious, we set $m \omega
  =1$ so that the mapping coordinates and momenta are defined to have
  dimensions of $\sqrt{\hbar}$.}
\bibitem[{foo({\natexlab{b}})}]{footnote1}
\bibinfo{note}{This set of evolution equations appears in
  Ref.~[\onlinecite{ThoSto99}] where a mapping description of the dynamics was
  investigated in the context of a semiclassical formulation.}
\bibitem[{foo({\natexlab{c}})}]{footnote2}
\bibinfo{note}{Efficient algorithms can be constructed to simulate these
  equations~\cite{kelly10} that account for the time scale disparity between
  the classical and mapping quantum degrees of freedom.}
\bibitem[{foo({\natexlab{d}})}]{footnote3}
\bibinfo{note}{Depending on the nature of the potential and the choice of
  initial conditions, there are situations where the dynamics described by this
  set of equations will yield unphysical results unless modifications to the
  dynamics are introduced~\cite{stotho,bonella01,kelly10}.}
\end{thebibliography}
\end{document}